# PROGRESS ON THE JOURNEY TO PUT FIELD ELECTRON EMISSION ONTO A BETTER SCIENTIFIC BASIS

*Richard G. Forbes[1], Sergey V. Filippov[2], Anatoly G. Kolosko[2], and Eugeni O. Popov[2]*
[1]University of Surrey, Advanced Technology Institute & School of Computer Science and Electronic Engineering, Guildford, Surrey GU2 7XH, UK
[2]Ioffe Institute, Division of Plasma Physics, Atomic Physics and Astrophysics, St Petersburg 194021, Russia

## ABSTRACT
This presentation is part of a long-term project to put field electron emission (FE) onto a better scientific basis, by seeking reliable quantitative agreement between theory and experiment, especially as regards emission-current values.

The main paper aims are: (1) to respond to remarks made in recent papers [1,2]; (2) to restate the thinking behind our 2022 methodology [3] for choosing between different FE models using experiments; (3) to assess progress; and (4) to make further suggestions about improved approaches.

## KEYWORDS
Field electron emission; theory-experiment comparison; AHFP exponent $\kappa$; decision table analysis.

## BACKGROUND
An *electronically ideal* FE system has its measured current-voltage [$I(V)$] characteristics determined by the emission physics and zero-current system electrostatics alone. Many modern FE systems are not ideal [4]. Basic scientific experiments need to be done on systems that are as simple as practicable; thus, discussion here relates only to electronically ideal emitters/systems that behave as good conductors.

Major historical FE theories are the 1928/29 theory of Fowler and Nordheim (FN) [5], based on an exactly triangular (ET) tunneling barrier, and the 1956 theory of Murphy and Good (MG) [6], based on the Schottky-Nordheim (SN) barrier. MG theory corrected errors found in earlier work (see [7]). It is *theoretically certain* that, although MG theory is not "fully correct physics", it is "better physics" than FN's. A valid methodology of experiment-based theory-comparison ought to be able to show this, and establishing this capability is a first step.

In 2008 [8], RGF suggested that comparisons could be based on the mathematical form (now described [3] as the "AHFP current-voltage equation")

$$I = CV^{\kappa}\exp[-B/V], \quad (1)$$

where $B$ and $C$ are constants, and the so-called *AHFP exponent* $\kappa$ is theory-based. Form (1) can represent both ET and SN barrier physics. ET barrier physics yields $\kappa=2$; SN barrier physics yields the work-function ($\phi$) related value

$$\kappa \approx 2 - [1.64 \, (\text{eV}/\phi)^{1/2}]. \quad (2)$$

If $\phi = 4.50$ eV then $\kappa \approx 1.23$, so in 2008 it looked as if accurate measurement of $\kappa$ could decide between theories. But by 2022 we knew (see [3]) that other factors also influenced $\kappa$, that the FN and MG theories yielded different allowable *ranges* of $\kappa$, and that the methodology in its 2022 form (although looking promising) did not yield a decisive result.

## COMMENT ON A RECENT PAPER [1]
Although eq. (2) gives an initial impression that an experimental $\kappa$-value could yield an experimental $\phi$-value, the present authors have *never* considered that eq. (2) could be used reliably in this way. In 2008 experimental data were too noisy to get a reliable $\kappa$-value; by 2022 we knew that other factors significantly contribute to measured $\kappa$-values.

Ayari et al. [1] have explored in numerical detail whether in principle eq. (2) (or other FE methods) could be used to extract a value of $\phi$ from experimental FE data. They concluded that the practical difficulties are too great. This accords with our view, and it is useful to have the numerics set out in detail. They then go on to suggest that eqns (1) and (2) have limited scientific use. For finding $\phi$ we agree; but this is not what we are using these equations for. We also stress that we are treating eq. (1) primarily as a mathematical form, not as an approximated version of the MG FE equation (which it also is).

## PRINCIPLES BEHIND OUR 2022 WORK [3]
### The use of an "intermediate-status" approach
In FE, for currents, direct theory-experiment comparisons have never proved decisive, due mainly to the large uncertainty associated with theoretical predictions. Our 2022 alternative approach [3] derives from chemistry, where "intermediate-status" parameters (such as free energies) are defined, and can be derived both theoretically and experimentally. A form of this approach was successful in choosing between models in field evaporation theory [9]. Equation (1) is an "intermediate-status" equation where $\kappa$ can be estimated both experimentally and theoretically. It is not the only possible equation (see [3]), but it seems the simplest and hence deserves to be explored first. We emphasize that our current work is an "exploration in progress".

### Estimation of FN and MG $\kappa$-ranges
For electronically ideal FE systems, ref. [3] identified six factors able to affect the value of $\kappa$, as follows. A: emitter band-structure. B: barrier form (i.e., barrier shape). C: the atomic-level wave-functions of surface atoms. D: voltage dependence in the notional emission area (as affected by emitter shape and the distribution of local work function). E:

other physical factors related to uncertainties in prediction of local current density. F: other uncertainties relating to the prediction of emission currents.

Table 1 is a slightly modified version of our original estimates [3] of the ranges of predicted $\kappa$-values associated with each effect, for each of the ET and SN barriers. (The range for D is now assessed differently.) Changes are shown bold. Table 2 shows the resulting decision table [3]; changes are shown in blue. As in [3], the only experimental result considered usable ($\kappa_m$=1.65) proves indecisive.

*Table 1: Ranges table, modified from ref. [3].*

| Table 1 (needle/post) sources of $\Delta\kappa$: | FN (ET barrier) | | MG (SN barrier) | |
|---|---|---|---|---|
| | lower limit | upper limit | lower limit | upper limit |
| A: | 2 | 2 | 2 | 2 |
| B: | 0 | 0 | − 0.77 | − 0.77 |
| C: | − 1 | 0 | −1 | 0 |
| D: | 0.58 | 1.03 | **0.62** | **1.06** |
| E: | ~0 | ~0 | ~0 | ~0 |
| F: | not known | not known | not known | not known |
| Range of $\kappa$ | 1.58 | 3.03 | **0.85** | **2.29** |

*Table 2: Decision table, modified from ref. [3].*

| Table 2: (needle/post) | Decision Ranges | Deduction [based on revised definition of limits] |
|---|---|---|
| X: | $\kappa_m$ < 0.85 | Not compatible with either theory |
| MG56: | 0.85 ≤ $\kappa_m$ < 1.58 | 1956 MG theory indicated |
| ? | 1.58 ≤ $\kappa_m$ ≤ 2.29 | INDECISIVE: compatible with MG or with FN |
| FN29: | 2.29 < $\kappa_m$ ≤ 3.03 | 1928/29 FN theory indicated |
| X: | 3.03 < $\kappa_m$ | Not compatible with either theory |

## NEW SIMULATION RESULTS

As described in [10], we have now extended our SN-barrier simulations to cover a wider range of emitter shapes and assumed work-functions. This results in revised Tables (3 and 4, below). We have assumed that our new simulations allow us to eliminate or disregard uncertainties of type F.

Outcomes are some slight changes in the ranges in the decision table, but no change in the conclusion that the experimental result ($\kappa_m$=1.65) proves indecisive. It is obvious, without detailed simulations, that wider exploration of ET barrier effects would not change this conclusion.

## DISCUSSION
**Desirability of well-defined flat-surface experiments**

In overall terms, it is clear that what needs to be done is to reduce the sources of uncertainty in the decision table. Experiments on well-defined flat surfaces (rather than pointed emitters) would eliminate uncertainties of type D and result in the following "flat-surface" tables (5 and 6).

*Table 3: Ranges table, using revised type-D values.*

| Table 3 (needle/post) sources of $\Delta\kappa$: | FN (ET barrier) | | MG (SN barrier) | |
|---|---|---|---|---|
| | lower limit | upper limit | lower limit | upper limit |
| A: | 2 | 2 | 2 | 2 |
| B: | 0 | 0 | − 0.77 | − 0.77 |
| C: | − 1 | 0 | −1 | 0 |
| D: | 0.58 | 1.03 | **0.57** | **1.07** |
| E: | ~0 | ~0 | ~0 | ~0 |
| Range of $\kappa$ | 1.58 | 3.03 | **0.90** | **2.30** |

*Table 4: Decision table, using revised type-D values.*

| Table 4: (needle/post) | Decision Ranges | Deduction [based on revised values of type D] |
|---|---|---|
| X: | $\kappa_m$ < 0.90 | Not compatible with either theory |
| MG56: | 0.90 ≤ $\kappa_m$ < 1.58 | 1956 MG theory indicated |
| ? | 1.58 ≤ $\kappa_m$ ≤ 2.30 | INDECISIVE: compatible with MG or with FN |
| FN29: | 2.30 < $\kappa_m$ ≤ 3.03 | 1928/29 FN theory indicated |
| X: | 3.03 < $\kappa_m$ | Not compatible with either theory |

*Table 5: Ranges table for flat emitter surface.*

| Table 5 (flat surface) sources of $\Delta\kappa$: | FN (ET barrier) | | MG (SN barrier) | |
|---|---|---|---|---|
| | lower limit | upper limit | lower limit | upper limit |
| A: | 2 | 2 | 2 | 2 |
| B: | 0 | 0 | − 0.77 | − 0.77 |
| C: | − 1 | 0 | −1 | 0 |
| E: | ~0 | ~0 | ~0 | ~0 |
| Range of $\kappa$ | 1.00 | 2.00 | 0.23 | 1.23 |

*Table 6: Decision table for flat emitter surface.*

| Table 6: (flat surface) | Decision Ranges | Deduction [based on flat surfaces of known area and local work function 4.50 eV] |
|---|---|---|
| X: | $\kappa_m$ < 0.23 | Not compatible with either theory |
| MG56: | 0.23 ≤ $\kappa_m$ < 1.00 | 1956 MG theory indicated |
| ? | 1.00 ≤ $\kappa_m$ ≤ 1.23 | INDECISIVE: compatible with MG or with FN |
| FN29: | 1.23 < $\kappa_m$ ≤ 2.00 | 1928/29 FN theory indicated |
| X: | 2.00 < $\kappa_m$ | Not compatible with either theory |

With Table 6 the experimental result $\kappa_m$=1.65 (which is for a carbon-nanotube (CNT) large area field emitter—see [3]) at first sight seems to indicate that 1928/29 FN theory is applicable. However, a more plausible interpretation is that effects of type D are "pushing up" the value of $\kappa_m$ for a CNT, and that the need is for reliable flat-surface experiments. An earlier (noisy) analysis of tungsten results (see Fig. 1) showed no evidence that $\kappa_m$ was significantly different from zero.

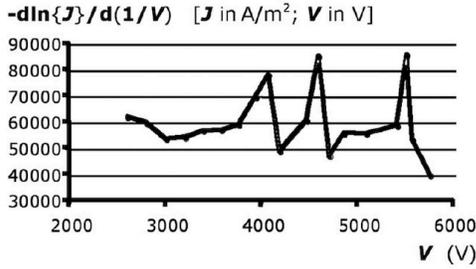

*Figure 1: Noisy experimental data plot for tungsten emitter, reproduced from Fig. 2 in [8]. Theoretically the plot should be smooth, with slope equal to $\kappa_m$ (see [8]).*

**Experimental flat-surface measurements**

At present, probably the easiest way of conducting a "nearly flat surface" experiment is to use a smooth metal emitter of large apex radius. Two experimental configurations seem of interest. The first is to take measurements from a suitably placed probe-hole, using traditional field electron microscope geometry. Phase-sensitive detection and electronic recording of current-voltage measurements would need to be used.

A second approach is to use a modified (reverse-biased) atom probe with a probe-hole set on the interior of a flat crystal facet of known crystallographic orientation. With this approach one could determine areas by counting atoms. A difficulty would be how to deal with the variation of the surface electrostatic field across the area of observation, but a merit is that atom probes operate under good ultra-high-vacuum conditions. This approach merits careful exploration. An earlier version [11] gave promising results.

**Noise and sensitivity issues**

This paper has focused on uncertainties related to theoretical prediction of the AHFP exponent $\kappa$. There are also uncertainties associated with reliable determination of an experimental value $\kappa_m$, not least if circumstances are such that field dependences are involved. (However, problems of this kind would be reduced in a flat-surface experiment.)

Modern data-analysis methodologies for extracting $\kappa_m$-values were discussed in [12] and also in [1] and [2]. However, it seems preferable to leave further discussion to papers that discuss the detailed design of apparatus. We note, though, that "noise" issues relating to the extraction of $\kappa_m$-values seem to be significantly less sensitive than those relating to extracting $\phi$-values.

**Need to explore surface-atom-related effects**

It remains to be discovered whether flat-surface experiments will enable a decisive experimental choice between 1929 FN theory and 1956 MG theory. An alternative is to attempt to reduce the uncertainty range associated with surface-atom-related (type C) effects. This is expected to be difficult but needs to be pursued.

**CONCLUSIONS**

Extended simulations of shape and work-function effects have not enabled a decisive experimental choice between 1929 FN theory and 1956 MG theory. Next steps are to explore how to do experiments on flat surfaces and/or develop improved theory for surface-atom-related effects.

**REFERENCES**

[1] A. Ayari, P. Vincent, S. Perisanu, P. Poncharal, S.T. Purcell, "All field emission models are wrong, … but are any of them useful?", *J. Vac. Sci. Technol. B,* vol. 40, 024001, 2022.

[2] A. Ayari, P. Vincent, S. Perisanu, P. Poncharal, S.T. Purcell, "All field emission experiments are noisy, … are any of them meaningful?", *J. Vac. Sci. Technol. B*, vol. 41, 024001, 2023.

[3] S. V. Filippov, A. G. Kolosko, E. O. Popov, R. G. Forbes, "Field emission: calculations supporting a new methodology of comparing theory with experiment", *R. Soc. Open. Sci.*, vol. 9, 220748, 2022.

[4] R. G. Forbes, "Proposal that interpretation of field emission current-voltage characteristics should be treated as a specialized form of electrical engineering", *J. Vac. Sci. Technol. B,* vol. 41, 028501, 2023.

[5] T. E. Stern, B. S. Gossling, R. H. Fowler, "Further studies in the emission of electrons from cold metals", *Proc. R. Soc. Lond. A*, vol. 124, pp. 699-723, 1929.

[6] E. L. Murphy and R. H. Good, "Thermionic emission, field emission and the transition region", *Phys. Rev.*, vol. 192, pp. 1464-1473, 1956.

[7] R. G. Forbes, "Comments on the continuing widespread and unnecessary use of a defective equation in field emission related literature", *J. Appl. Phys.*, vol. 126, 210901 (2019).

[8] R. G. Forbes, "Call for experimental test of a revised mathematical form for empirical field emission current-voltage characteristics", *Appl. Phys. Lett.*, vol. 92, 193106, 2008.

[9] R. G. Forbes, R. K. Biswas, K. Chibane, "Field evaporation theory: re-analysis of published field-sensitivity data", *Surface Sci.* 114, pp. 498514, 1982.

[10] S. V. Filippov, R. G. Forbes, E. O. Popov, A. G. Kolosko, F. F. Dall'Agnol, "Further studies on using the AHFP exponent to choose between alternative field emission theories", Presentation, this conference.

[11] C. D. Ehrhlich and E. W. Plummer, "Measurement of the absolute tunneling current density in field emission from W(110)", Phys. Rev. B, vol. 18, p. 3767-3771, 1978.

[12] R. G. Forbes, E. O. Popov, A. G. Kolosko, S. V. Filippov, "The pre-exponential voltage-exponent as a sensitive test parameter for field emission theories", *R. Soc. Open Sci.* vol. 8, 201986, 2021.

**CONTACT**

\*R.G. Forbes: r.forbes@trinity.cantab.net